\documentclass[sigconf]{acmart}
\AtBeginDocument{%
  }

\setcopyright{acmlicensed}
\copyrightyear{2018}
\acmYear{2018}
\acmDOI{XXXXXXX.XXXXXXX}
\acmConference[Conference acronym 'XX]{Make sure to enter the correct
  conference title from your rights confirmation email}{June 03--05,
  2018}{Woodstock, NY}
\acmISBN{978-1-4503-XXXX-X/2018/06}

\usepackage{booktabs}
\usepackage{array}
\PassOptionsToPackage{table,xcdraw}{xcolor}  
\usepackage{xcolor}
\usepackage{colortbl}

\renewcommand\footnotetextcopyrightpermission[1]{} 
\newcounter{tightlistcnt}
\newenvironment{tightitemize}{%
	\begin{list}{\textbullet}{\usecounter{tightlistcnt}%
			\topsep 0in
			\partopsep 0in
			\itemsep 0in
			\parsep 0in
			\leftmargin 0em
			\rightmargin 0in
			\listparindent 0em
			\itemindent .75em
			\labelsep .75em
			\labelwidth 0in
		}%
	}{%
	\end{list}%
}

\newenvironment{tightenumerate}{%
	\begin{list}{\arabic{tightlistcnt}.}{\usecounter{tightlistcnt}
			\topsep 0in
			\partopsep 0in
			\itemsep .2em
			\parsep 0in
			\leftmargin 0.0em
			\rightmargin 0in
			\listparindent 0em
			\itemindent .75em
			\labelsep .75em
			\labelwidth 0in
		}%
	}{%
	\end{list}%
}

\begin{document}

\title{RAG for Geoscience: What We Expect, Gaps and Opportunities} 

\author{
	Runlong Yu$^{1}$, Shiyuan Luo$^{2}$, Rahul Ghosh$^{3}$, Lingyao Li$^{4}$, Yiqun Xie$^{5}$, Xiaowei Jia$^{2}$\\
        $^{1}$University of Alabama,
	$^{2}$University of Pittsburgh, 
	$^{3}$University of Minnesota, Twin Cities\\
	$^{4}$University of South Florida,
	$^{5}$University of Maryland, College Park\\
	 ryu5@ua.edu, \{shl298, xiaowei\}@pitt.edu, ghosh128@umn.edu, lingyaol@usf.edu, xie@umd.edu
}

\renewcommand{\shortauthors}{Runlong Yu et al.}

\begin{abstract}
Retrieval-Augmented Generation (RAG) enhances language models by combining retrieval with generation. However, its current workflow remains largely text-centric, limiting its applicability in geoscience. Many geoscientific tasks are inherently \emph{evidence-hungry}. Typical examples involve imputing missing observations using analog scenes, retrieving equations and parameters to calibrate models, geolocating field photos based on visual cues, or surfacing historical case studies to support policy analyses. A simple ``retrieve-then-generate'' pipeline is insufficient for these needs.
We envision \textit{Geo-RAG}, a next-generation paradigm that reimagines RAG as a modular \textit{retrieve $\rightarrow$ reason $\rightarrow$ generate $\rightarrow$ verify} loop. Geo-RAG supports four core capabilities: (i) retrieval of multi-modal Earth data; (ii) reasoning under physical and domain constraints; (iii) generation of science-grade artifacts; and (iv) verification of generated hypotheses against numerical models, ground measurements, and expert assessments. This shift opens new opportunities for more trustworthy and transparent geoscience workflows. 
\end{abstract}

\maketitle

\vspace{-0.15cm}
\section{Introduction}
\vspace{-0.05cm}

Retrieval-Augmented Generation (RAG) has emerged as a widely adopted paradigm to enhance large language models (LLMs) by integrating external knowledge during generation~\cite{lewis2020retrieval, izacard2022few, guu2020retrieval}. At its core, RAG retrieves relevant content from a knowledge corpus (e.g., documents, web resources) and feeds it into a language model to produce more informed and context-aware outputs. This simple yet powerful idea has led to notable successes in open-domain question answering~\cite{karpukhin2020dense, izacard2020distilling}, code completion~\cite{chen2021evaluating}, and domain-specific chatbot development~\cite{shi2023replug}. With the rising demand for trustworthy AI systems, RAG provides a compelling mechanism to ground generation in timely and verifiable knowledge sources, thereby mitigating hallucinations and improving factual accuracy~\cite{shi2023replug, lewis2020retrieval}.

\vspace{-0.17cm}
\subsection{Success Stories from RAG}
\subsubsection{Open-Domain Question Answering and Search Augmentation.}  
RAG has shown remarkable performance in knowledge-intensive tasks, particularly in open-domain question answering~\cite{lewis2020retrieval}. Systems like Google’s Search-Augmented Models and Meta’s original RAG framework combine retrieval from large corpora (e.g., Wikipedia) with generation, allowing models to answer questions that would otherwise exceed their parametric memory. This approach enables factual consistency while remaining flexible and scalable, and has been widely adopted in digital assistants and enterprise search~\cite{karpukhin2020dense, guu2020retrieval}.

\vspace{-0.1cm}
\subsubsection{Domain Adaptation and Low-Resource Knowledge Access.}  
In specialized domains such as biomedicine, law, and customer service, RAG has been used to inject domain-specific corpora into the generation process, offering low-cost adaptation without full retraining~\cite{deyoung2021ms2, wu2024retrieval}. For instance, medical QA systems can retrieve research articles or clinical guidelines to support diagnoses; legal assistants can cite relevant regulations in generated summaries.

\vspace{-0.2cm}
\subsection{Why RAG for Geoscience Is Different}

Geoscientific questions rarely reduce to simply ``finding a passage and summarizing it.'' Instead, they involve \emph{evidence-hungry} tasks, where models should go beyond parametric memory and surface-level language generation. Effective reasoning in these contexts requires retrieving, citing, and grounding outputs in external, structured, and domain-specific knowledge~\cite{karniadakis2021physics, xie2023geo}. Typical examples include imputing missing observations using analog scenes, retrieving equations and parameters to calibrate models, geolocating field photos based on visual cues, and surfacing historical case studies to support policy analyses.

Mainstream RAG cannot meet these needs: it takes in only text, lacks the awareness of underlying physical processes, emits format-agnostic outputs, and offers no built-in verification~\cite{lewis2020retrieval, khattab2022demonstrate, liu2023lost}. To address this gap, we introduce \textit{Geo-RAG}, a modular architecture built on a \textit{retrieve $\rightarrow$ reason $\rightarrow$ generate $\rightarrow$ verify} loop. It includes (i) retrieval of multi-modal Earth data; (ii) reasoning under physical and domain constraints; (iii) generation of science-grade artifacts; and (iv) verification of generated hypotheses against numerical models, ground measurements, and expert assessments. 

This paper makes three key contributions toward advancing RAG for geoscience:  
(i)~\textit{Problem reframing:} We formalize geoscientific tasks as evidence-hungry challenges and identify four representative scenarios that highlight fundamental gaps in mainstream RAG pipelines.  
(ii)~\textit{Geo-RAG blueprint:} We introduce \textit{Geo-RAG}, a modular architecture that integrates multimodal retrieval, physics-aware reasoning, structured artifact generation, and automated verification into a self-refining loop.  
(iii)~\textit{Research agenda:} We surface five core implementation challenges and outline governance issues that should be addressed to ensure Geo-RAG operates safely, equitably, and credibly in high-stakes geoscience settings.


\begin{table*}[t]\small
	\centering
	\caption{Core geoscience use cases and how \textit{Geo-RAG} advances them through the
		\textbf{retrieve $\rightarrow$ reason $\rightarrow$ generate $\rightarrow$ verify} loop.}
	\vspace{-0.35cm}
	\label{tab:geo_rag_scenarios}
	\rowcolors{2}{gray!10}{white}
	\begin{tabular}{p{2.4cm} p{3.3cm} p{6.6cm} p{4.1cm}}
		\toprule
		\textbf{Scenario} & \textbf{Illustrative Goal} & \textbf{Geo-RAG Contribution} & \textbf{Open Gaps (mainstream RAG)} \\
		\midrule
		Filling data gaps in Earth observations & 
		Data completion from nearby observations or historical analogs; reinforce simulations in sparse regions &
		\textbf{Retrieve:} multi-modal analog scenes and simulation priors.
		\textbf{Reason:} fuse with physics-aware constraints for spatial/temporal consistency.
		\textbf{Generate:} gap-filled rasters or ensemble forecasts with inline uncertainty.
		\textbf{Verify:} cross-validate against independent sensors, propagate error bars. &
		No raster/time-series indexing; blind to spatio-temporal co-registration and physical laws; zero uncertainty propagation. \\
		
		Scientific knowledge retrieval &
		Auto-assemble governing equations, parameters, and boundary conditions &
		\textbf{Retrieve:} symbolic formulae, parameter tables, legacy code.
		\textbf{Reason:} dimensional checks, context pruning, unit harmonisation.
		\textbf{Generate:} executable snippets / config files for simulators.
		\textbf{Verify:} provenance citations, unit tests. &
		Symbolic math unindexed; lacks unit/dimension validation; outputs prose only, no runnable configs or code. \\
		
		Spatial reasoning from visual inputs &
		Geolocate field photos, UAV videos, disaster snapshots &
		\textbf{Retrieve:} visually and contextually similar scenes, DEMs, metadata.
		\textbf{Reason:} global spatial reasoning under physics \& terrain constraints.
		\textbf{Generate:} probabilistic geolocation maps + geo-aware text.
		\textbf{Verify:} confidence calibration via auxiliary sensors / crowdsourcing. &
		Image \& DEM retrieval absent; no terrain-aware spatial reasoning; produces single uncalibrated textual guess. \\
		
		Case-based scientific decision support &
		Generate counterfactual scenarios and policy options &
		\textbf{Retrieve:} analogous historical cases, simulation archives, policy docs.
		\textbf{Reason:} causal-graph reasoning for ``what-if’’ analysis.
		\textbf{Generate:} counterfactual trajectories, impact metrics, policy briefs.
		\textbf{Verify:} simulate outcomes, enforce domain constraints, ensure traceability. &
		Cannot surface cross-modal causal cases; no counterfactual simulation loop; provenance \& fairness auditing absent. \\
		\bottomrule
	\end{tabular}
	\vspace{-0.43cm}
\end{table*}

\vspace{-0.2cm}
\section{Motivating Scenarios} \label{sec:2}

Here, we present four motivating scenarios that highlight the unique demands of geoscientific tasks. 

\textit{Scenario 1: Filling Data Gaps in Earth Observations.} 
Earth observations are frequently sparse or incomplete,  
especially in remote areas, polar environments, or time periods with limited historical coverage. This data gap is often caused by logistical challenges, harsh conditions, and sensor limitations. Moreover, cloud cover, instrument failures, and the high cost of field data collection can further lead to missing or incomplete records~\cite{jensen2009remote, ma2019deep, zhang2016deep}. Addressing these limitations---for example, by reconstructing surface temperature in mountainous terrain or filling gaps in satellite imagery requires integrating information from nearby observations, historical analogs, and physics-based simulations.

\textit{Scenario 2: Retrieving Scientific Knowledge for Modeling.}
Geoscientific modeling requires domain-specific knowledge, such as empirical formulas, parameter values, and boundary conditions from scientific literature or legacy code~\cite{gil2007examining,10.1145/3725982}. These tasks demand symbolic reasoning that spans across textual descriptions, mathematical expressions, and domain-specific rules.

\textit{Scenario 3: Spatial Reasoning from Visual Inputs.}
Geoscientists analyze field photographs or UAV-captured videos to identify geographic locations, landforms, or geohazard indicators~\cite{bischke2019multi, xia2018dota}. This involves retrieving visually and contextually similar scenes and reasoning over topographic, climatic, and geomorphological features. 

\textit{Scenario 4: Case-Based Scientific Decision Support.}
Scientific decisions in areas such as land use planning, disaster mitigation, or climate adaptation often rely on reasoning from past cases. Retrieving analogous case studies, simulation results, and policy documents enables counterfactual analysis, such as exploring the impacts of increased rainfall or rising temperatures under similar conditions.

\vspace{-0.2cm}
\section{What’s Missing and Why It Matters} \label{sec:3}

Table~\ref{tab:geo_rag_scenarios} outlines four representative scenarios and the key capabilities required to support them. While mainstream RAG has proven effective in text-based question answering, it falls short when applied to geoscientific tasks. These shortcomings manifest along four axes: it takes in only text, lacks physics awareness, emits format-agnostic outputs, and offers no built-in verification. These missing components directly limit the reliability, scalability, and trustworthiness of AI systems in geoscience. We unpack each axis below. 

\textit{Limitation 1: Text-Only Retrieval.}  
Mainstream RAG operates exclusively on unstructured text, whereas geoscientific reasoning relies on diverse data types—including satellite imagery, reanalysis datasets, seismic waveforms, and hydrological graphs. The process of ``flattening'' such structured data into text discards critical information, including spatial topology and numerical integrity, which degrades downstream reasoning. Moreover, text embedding retrieval methods (e.g., sentence-BERT~\cite{reimers2019sentence}) that ignore physical units and coordinate frames often surface semantically similar but physically invalid content. At scale, tokenizing petabyte-scale datasets becomes computationally prohibitive, making text-only architectures impractical for real-world geoscience scenarios.

\textit{Limitation 2: Physics-Blind Reasoning.}  
LLMs lack an inherent encode of physical laws such as conservation of mass or energy, and do not enforce dimensional consistency. As a result, they may produce physically implausible outputs—for example, negative rainfall or ocean temperatures exceeding 1000°C. These errors are not cosmetic; they can destabilize simulations, misguide operational forecasts, and undermine expert confidence~\cite{yu2025foundation}. Without reasoning grounded in physical constraints, RAG outputs cannot reliably support modeling or decision-making in scientific domains.

\textit{Limitation 3: Format-Insensitive Generation.}  
Mainstream RAG typically produces generic text outputs, a format that is often misaligned with structured formats and specialized terminology required in geoscience. In practice, domain experts do not need paragraphs; they need science-grade artifacts: gap-filled NetCDF rasters, executable model configuration files, spatially explicit geolocation maps, or regulatory-ready policy briefs enriched with geospatial semantics~\cite{janee2009preserving}. Hence, current RAG outputs are not compatible with these GIS tools, simulators, or operational systems. This forces scientists to manually post-process the data, a slow and error-prone process that can create a major bottleneck in their work. 

\textit{Limitation 4: Verification-Free Outputs.}  
Most RAG pipelines terminate after generation, with no assessment of output reliability. They provide no uncertainty estimates, no cross-sensor consistency checks, and no traceable links or evidence attribution to source data. This lack of verifiability is particularly problematic in high-stakes contexts such as disaster response or climate policy, where accountability and reproducibility are essential. For example, a flood risk map without confidence intervals may mislead evacuation planning, 
while unverifiable claims can complicate scientific audits and regulatory approval. Without verification, RAG outputs cannot meet the evidentiary standards required in science and policy.

\begin{figure}[t]
	\centering
	\includegraphics[width=0.84\linewidth]{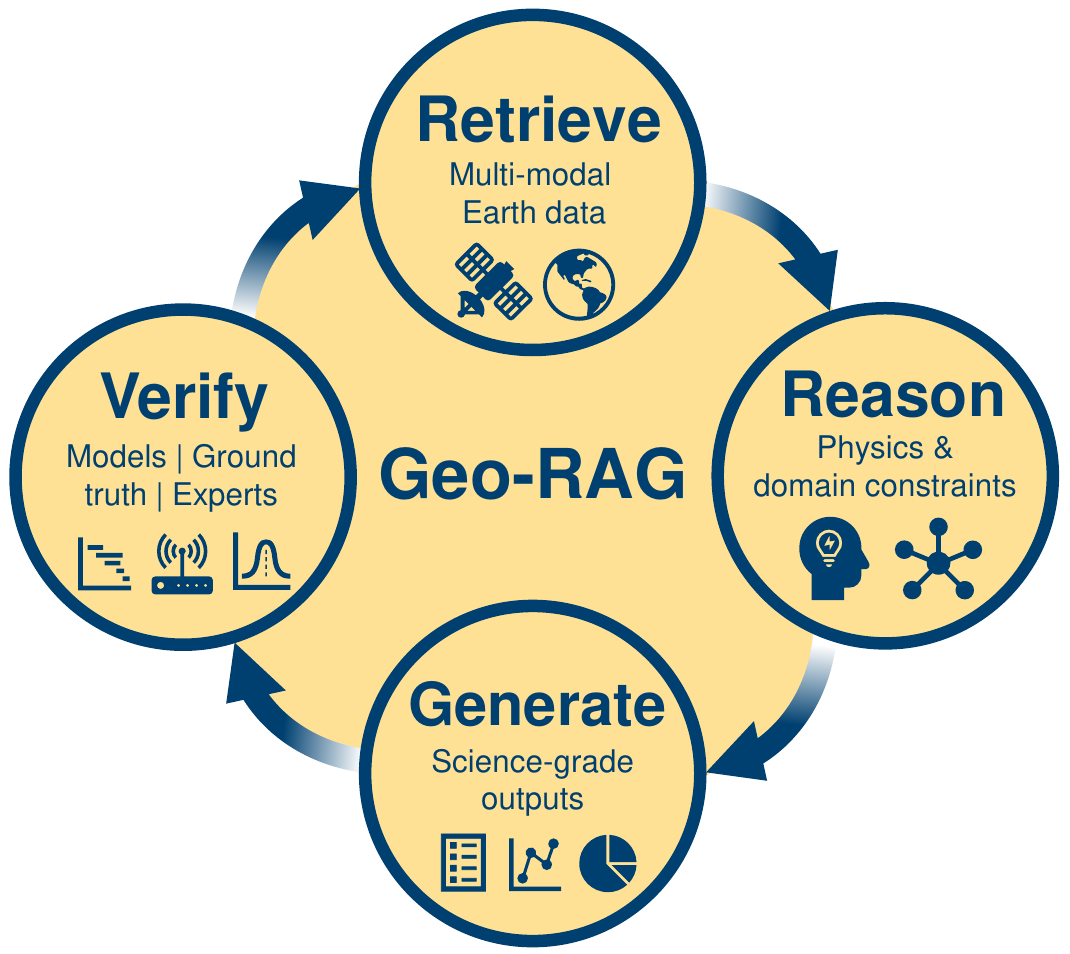}
		\vspace{-0.25cm}
	\caption{The \textit{retrieve $\rightarrow$ reason $\rightarrow$ generate $\rightarrow$ verify} loop.}
	\vspace{-0.35cm}
	\label{fig:geo_rag_framework}
\end{figure}

\vspace{-0.2cm}
\section{Vision of Geo-RAG} \label{sec:4}

Geo-RAG reimagines mainstream RAG as a four-stage, self-refining loop (Fig.~\ref{fig:geo_rag_framework}). By feeding validated evidence back into subsequent retrieval cycles, it transforms generation into an iterative, science-driven workflow.

\vspace{-0.2cm}
\subsection{Retrieve: Multi-Modal Earth Data}

Geo-RAG treats Earth itself as the primary knowledge base. Retrieval thus becomes a science-first process of surfacing evidence that is spatially anchored, temporally precise, physically interpretable, and structurally eclectic. This includes a set of artifacts: Sentinel-2 scenes~\cite{drusch2012sentinel}, weather-station time series, digital-elevation-model (DEM) tiles, reanalysis cubes~\cite{hersbach2020era5}, symbolic equations, and even dusty FORTRAN outputs. Across our motivating scenarios, Geo-RAG retrieves artifacts that co-register in the real world: thermal swaths and gauge data that overlap in space-time, calibration tables buried in legacy reports, DEM tiles aligned with UAV imagery, and case archives poised for counterfactual replay. For instance, a query may retrieve MODIS thermal swaths corresponding to temperature anomalies in wildfire-affected regions, matched with in-situ weather station data within a specified time window. This is driven by a hybrid indexing architecture: dense CLIP-style embeddings for language and imagery~\cite{radford2021learning}, geohash or Hilbert curves to encode spatial locality~\cite{sagan2012space}, balanced B-trees for temporal access, and abstract-syntax-tree (AST) hashing for symbolic code and equations~\cite{allamanis2018survey}. A multi-modal query planner decomposes complex questions and reassembles hits using physics-aware re-ranking, while unit-aware embeddings~\cite{griffioen2019unit} ensure that retrieved quantities preserve their physical semantics. In contrast to mainstream RAG, Geo-RAG preserves the numerical, spatial, and semantic integrity that is important for reasoning, generation, and verification.

\vspace{-0.15cm}
\subsection{Reason: Physics and Domain Constraints}

Once the right evidence is in hand, Geo-RAG reasons not in token space, but in the metric space of science, where every inference should obey conservation laws, dimensional coherence, causal ordering, and domain-specific ontologies. 
For example, precipitation infills are constrained by watershed water balance; snow–albedo feedback estimates are bounded by known physical ranges (e.g., albedo $\in [0, 1]$); parameter sets that fail Buckingham-$\pi$ dimensional checks~\cite{barenblatt1996scaling} are rejected; elevation candidates inconsistent with solar geometry or slope profiles are pruned; and counterfactual simulations remain bounded within empirically calibrated climate sensitivities.
This scientific rigor is operationalized through a neuro-symbolic reasoning stack: a dimensional sentry flags unit mismatches; differentiable constraint solvers encode conservation laws as soft penalties~\cite{yu2024adaptive}; and semantic filters built on Climate and Forecast (CF) conventions~\cite{eaton2003netcdf} and World Meteorological Organization (WMO) codes~\cite{wmo2011manual} eliminate structurally implausible paths. Instead of selecting the most likely continuation, Geo-RAG filters and ranks hypotheses based on their physical admissibility, each annotated with a transparent ``physics audit trail'' that explains why it survived. The output of this process is scientifically grounded scaffolding ready to inform structured generation.

\vspace{-0.25cm}
\subsection{Generate: Science-Grade Artifacts}

Geo-RAG elevates language models from narrative generators to producers of executable artifacts. Instead of producing generic prose, it generates artifacts that are directly usable within scientific workflows: CF-compliant NetCDF rasters with pixel-level uncertainty bands, simulation-ready configuration files for models like WRF~\cite{skamarock2008description}, probabilistic geolocation products packaged as GeoTIFF plus GeoJSON, and policy briefs enriched with causal diagrams and traceable references. These outputs are schema-aligned, standards-compliant, and machine-ingestible—designed to flow seamlessly into geospatial tools, numerical simulators, and policy pipelines without post-processing. 
To support this, we envision equipping Geo-RAG with a template-driven generation engine grounded in domain standards (e.g., CF conventions~\cite{eaton2003netcdf,bourhis2017json}), where templates are expressed as JSON-Schema objects~\cite{pezoa2016foundations}. Uncertainty quantification is propagated from upstream reasoning and formalized using techniques such as Monte Carlo dropout~\cite{gal2016dropout} and ensemble modeling~\cite{lakshminarayanan2017simple}. By aligning outputs with operational needs, Geo-RAG transforms language models from suggestive advisors into fully embedded components of scientific infrastructure.

\vspace{-0.25cm}
\subsection{Verify: Models, Ground Truth, and Experts}

A generation system without verification is not a scientific tool. Geo-RAG completes the loop by turning retrieval-augmented generation into a self-refining scientific instrument, where every output is treated as a testable hypothesis. We envision Geo-RAG implementing three concentric layers of validation. (1) Model replay injects generated artifacts (e.g., rasters or configuration files) into simulators, comparing outputs against withheld observational data using established metrics. (2) Sensor cross-checks evaluate consistency with independent sources, flagging discrepancies like spatial residuals in soil-moisture estimates that violate physical thresholds. (3) Expert-in-the-loop review enables scientists to interactively inspect, edit, and approve elements such as causal graphs or model namelists, with all feedback propagated upstream. 
This process is orchestrated by a containerized ensemble validation system that automates model injection, computes diagnostic scores, logs traceable outcomes, and refines uncertainty estimates. Each artifact is tagged with W3C PROV-O provenance metadata~\cite{w3c-prov-o} to ensure end-to-end auditability.
If verification fails, Geo-RAG automatically retrieves new evidence and re-enters the  \textit{retrieve $\rightarrow$ reason $\rightarrow$ generate} cycle. 


Together, these four stages position Geo-RAG as a blueprint for next-generation geoscientific reasoning systems, which are physically grounded, structurally validated, uncertainty-aware, and operationally usable. Geo-RAG elevates RAG into a full-loop scientific workflow, adaptable across domains from wildfire monitoring to hydrological forecasting and climate resilience planning.

\vspace{-0.2cm}
\section{Geo-RAG in Action}
To make Geo-RAG concrete, consider completing satellite imagery in cloud-covered, high-relief terrain, common in post-wildfire monitoring (revisiting \textit{Scenario 1}). Geo-RAG retrieves candidate artifacts (e.g., historical scenes, simulation priors, weather data) selected not for semantic similarity but for physical relevance, respecting spatial alignment, sensor characteristics, and temporal context. These candidates are then filtered through physics-aware reasoning. Artifacts inconsistent with energy balance, terrain geometry, or unit coherence are pruned, ensuring that only physically plausible evidence informs generation. The generation stage then constructs a structured output: for example, a CF-compliant NetCDF raster with inline uncertainty, suitable for downstream modeling or policy use. The generated artifact is treated as a hypothesis to be tested. Verification proceeds through model replay, sensor cross-checks, or expert review. If the artifact induces unrealistic dynamics or violates known constraints, this feedback triggers a return to retrieval, now informed by the nature of the failure. The system reselects evidence, adapts its reasoning, and iterates.

\vspace{-0.2cm}
\section{Implementation Challenges} \label{sec:5}

We now turn to the more practical question: How does Geo-RAG collide with low-level realities during construction? We outline five core technical tensions that emerge when transitioning Geo-RAG from principle to practice. These challenges reflect broader issues inherent to any attempt to operationalize scientific RAG at scale. 

Streaming index mutation presents a significant challenge. Standard approximate nearest neighbor (ANN) indices like HNSW or IVF assume static corpora~\cite{malkov2018efficient,jegou2010product}; however, Earth observations are continuously updated. As the data manifold shifts, recall degrades unless the index adapts accordingly. Dual-tier designs—with a stable base graph and a lightweight overlay refreshed hourly—provide a path forward, but the cost is steep: in-memory overlays can inflate resident memory by tens of gigabytes per day without aggressive eviction policies. The deeper issue lies not in indexing frequency, but in the lack of efficient structures that support mutable, temporally versioned, physics-aware retrieval.

A second tension arises in embedding unit-conditioned values. Tagging physical units during contrastive pretraining helps separate incompatible quantities like Celsius and Kelvin, but also warps the latent space, like cosine distance begins to reflect both semantic and unit affinity, confusing similarity metrics. Gradient instability occurs when conflicting units co-occur in the same batch. While mixup strategies can mitigate this issue, they often incur runtime overhead due to dynamic re-tokenization and on-the-fly unit conversions. Embeddings that respect unit semantics without sacrificing metric coherence remain elusive.

Constraint-aware decoding introduces a third bottleneck~\cite{liu2021co2sum}. To enforce physical laws such as mass or energy conservation, we embed soft constraints into the decoding process via a differentiable loss surface~\cite{yu2024adaptive,yu2025physics}. However, projecting hypotheses onto the constraint manifold during beam search dramatically increases latency, disrupting interactive use cases. A cascaded filtering approach, where a fast surrogate model pre-screens candidate completions before expensive optimization, helps reduce latency to a more tolerable level while preserving acceptable constraint satisfaction. Still, the tradeoff between responsiveness and rigor persists.

Generation of schema-conformant scientific artifacts reveals fragility. Structured outputs like NetCDF files should follow strict CF conventions in both field naming and dimensional alignment~\cite{eaton2003netcdf}. Lexical errors can silently crash downstream simulators, with no immediate indication of failure.  

Finally, the cost of verification forces difficult scheduling decisions. Running a full physics replay for every generated artifact is computationally prohibitive. To reduce overhead, we can employ an adaptive validator that escalates validation fidelity based on estimated uncertainty. Low-risk artifacts undergo lightweight sensor cross-checks; high-risk ones trigger full-model simulations. As real-world conditions shift with seasonal or climate-induced changes, the scheduler’s priors can become outdated, exposing the fragility of static validation heuristics.

\vspace{-0.2cm}
\section{Human Trust and Governance} \label{sec:6}
Trust and governance are foundational to any AI system deployed in the geosciences, where outputs frequently inform high-stakes decisions with implications for public safety, environmental stewardship, and geopolitical stability. Beyond technical performance, these systems should be transparent, equitable, and accountable, particularly in settings marked by sparse data, uneven infrastructure, or contested knowledge systems. Geo-RAG not only inherits these demands but amplifies them, as each stage of its architecture introduces distinct governance pressures. Retrieval should mitigate rich-data bias through equitable evidence selection; reasoning should uphold auditable, physics-informed constraints; generation and verification should maintain traceability and scientific validity. 

\vspace{-0.2cm}
\section{Conclusion} \label{sec:7}

Geoscience stands at a turning point. As environmental systems grow more complex and the stakes of decision-making rise, our tools for understanding the Earth should evolve from static models and siloed data to dynamic, integrated reasoning systems. Retrieval-Augmented Generation marks an important step in this direction, but its current incarnation remains too narrow, blind to physical laws, fragile under multimodal evidence, and unaccountable in high-consequence domains. Geo-RAG offers a new foundation: a modular, physics-grounded, and verification-aware architecture that reimagines RAG as a scientific workflow. By elevating retrieval to a spatial and temporal act of measurement, by embedding domain constraints into inference, and by treating each output as a falsifiable hypothesis, Geo-RAG aligns generative AI with the epistemic rigor of geoscience. This is not merely a technical proposal—it is a disciplinary vision. To move from data to decision, from prediction to understanding, geoscience should lead the way in shaping AI that is trustworthy, transparent, and deeply grounded in the structure of the natural world. Geo-RAG is a step toward that future: not just a system, but a call to reimagine how we build, validate, and govern scientific knowledge in the age of AI.

\bibliographystyle{ACM-Reference-Format}
\bibliography{mybibliography,survey}

\end{document}